\documentclass[conference,10pt,letterpaper]{IEEEtran}
\usepackage{amsfonts}

\newcommand{\set}[1]{\left\{ #1\right\}}
\newcommand{\gilt}{:}
\newcommand{\sodass}{\,:\,}
\newcommand{\setGilt}[2]{\left\{ #1\sodass #2\right\}}

\newcommand{\realrange}[2]{\left[#1, #2\right]}

\newcommand{\unitrange}[2]{\realrange{0}{1}}

\newcommand{\Oh}[1]{\mathcal{O}\!\left( #1\right)}

\newcommand{\llabel}[1]{\label{\labelprefix:#1}}
\newcommand{\labelprefix}{} 

\newcommand{\discussionsize}{\small}

\newcommand{\frage}[1]{}

\newenvironment{code}{\noindent
\begin{tabbing}%
\hspace{2em}\=\hspace{2em}\=\hspace{2em}\=\hspace{2em}\=\hspace{2em}\=%
\hspace{2em}\=\hspace{2em}\=\hspace{2em}\=\hspace{2em}\=\hspace{2em}\=%
\kill}{\end{tabbing}}

\newcommand{\labelcommand}{}
\newcommand{\captiontext}{}
\newsavebox{\codeparam}
\newcounter{lineNumber}
\newenvironment{disscodepos}[3]{%
\renewcommand{\labelcommand}{#2}%
\renewcommand{\captiontext}{#3}%
\sbox{\codeparam}{\parbox{\textwidth}{#3}}%
\begin{figure}[#1]\begin{center}\begin{code}\setcounter{lineNumber}{1}}{%
\end{code}\end{center}\caption{\llabel{\labelcommand}\captiontext}\end{figure}}

{\end{disscodepos}}

\newcommand{\Is}       {:=}

\newdimen\endofsize\endofsize=0.5em
\def\endofbeweis{~\quad\hglue\hsize minus\hsize
                 \hbox{\vrule height \endofsize width
\endofsize}\par}

\usepackage[utf8]{inputenc}
\usepackage{algorithmic}
\usepackage{algorithm}
\usepackage{amsmath}
\usepackage{amssymb}
\usepackage{color}
\usepackage{latexsym}
\usepackage{makeidx}
\usepackage{multicol}
\usepackage{numprint}
\usepackage{t1enc}
\usepackage{graphicx}
\usepackage{url}
\npdecimalsign{.} 

\definecolor{mygrey}{gray}{0.75}
\newcommand{\ie}{i.e.\ }
\newcommand{\etal}{et~al.\ }
\newcommand{\eg}{e.g.\ }
\newcommand{\Id}[1]{\ensuremath{\text{{\sf #1}}}}

\def\MdR{\ensuremath{\mathbb{R}}}

\newcommand{\mytitle}{Parallel Graph Partitioning for Complex Networks }
\begin{document}
\title{\mytitle}

\author{ \IEEEauthorblockN{Henning Meyerhenke}
   \IEEEauthorblockA{Karlsruhe Institute of Technology (KIT) \\
        Karlsruhe, Germany \\
        meyerhenke@kit.edu} 
   \and
 \IEEEauthorblockN{Peter Sanders}
  \IEEEauthorblockA{Karlsruhe Institute of Technology (KIT) \\
        Karlsruhe, Germany \\
        sanders@kit.edu} 

   \and
\IEEEauthorblockN{Christian Schulz} 
  \IEEEauthorblockA{Karlsruhe Institute of Technology (KIT) \\
        Karlsruhe, Germany \\
        christian.schulz@kit.edu}}
\date{11 April 2014}

\pagestyle{plain}

\maketitle
\begin{abstract}
  Processing large complex networks like social networks or web graphs has   recently attracted considerable interest. In order to do this in parallel, we   need to partition them into pieces of about equal size.  Unfortunately,   previous parallel graph partitioners originally developed for more regular   mesh-like networks do not work well for these networks.  This paper addresses   this problem by parallelizing and adapting the \emph{label propagation}   technique originally developed for graph clustering.  By introducing size   constraints, label propagation becomes applicable for both the coarsening and   the refinement phase of multilevel graph partitioning. We obtain very high   quality by applying a highly parallel evolutionary algorithm to the coarsened   graph. The resulting system is both more scalable and achieves higher quality   than state-of-the-art systems like ParMetis or PT-Scotch. For large complex   networks the performance differences are very big.  For example, our algorithm   can partition a web graph with 3.3 billion edges in less than \emph{sixteen     seconds} using 512 cores of a high performance cluster while producing a   high quality partition -- none of the competing systems can handle this graph on our system.
\end{abstract}

\IEEEpeerreviewmaketitle
\section{Introduction}
Graph partitioning (GP) is a key prerequisite for efficient large-scale parallel graph algorithms. 
A prominent example is the PageRank algorithm \cite{BrinP98}, which is used by search engines such as Google to order web pages displayed to the user by their importance. 
As huge networks become abundant, there is a need for their parallel analysis.
In many cases, a graph needs to be partitioned or clustered such that there are few edges between the blocks (pieces).  
In particular, when you process a graph in parallel on $k$ PEs (processing elements), you often want to partition the graph into $k$ blocks of about equal
size.  In this paper we focus on a version of the problem that constrains the
maximum block size to $(1+\epsilon)$ times the average block size and tries to
minimize the total cut size, i.e., the number of edges that run between blocks.

It is well-known that there are more realistic (and more complicated) objective
functions involving also the block that is worst and the number of its
neighboring nodes \cite{HendricksonK00}, but minimizing the cut size has been adopted as
a kind of standard since it is usually highly correlated with the other
formulations. 
The graph partitioning problem is NP-complete \cite{Hyafil73,Garey1974} and there is no approximation algorithm with a constant ratio factor for general graphs \cite{BuiJ92}. 
Hence, heuristic algorithms are used in practice.  

\begin{figure}[h]
\centering
\includegraphics[width=0.45\textwidth]{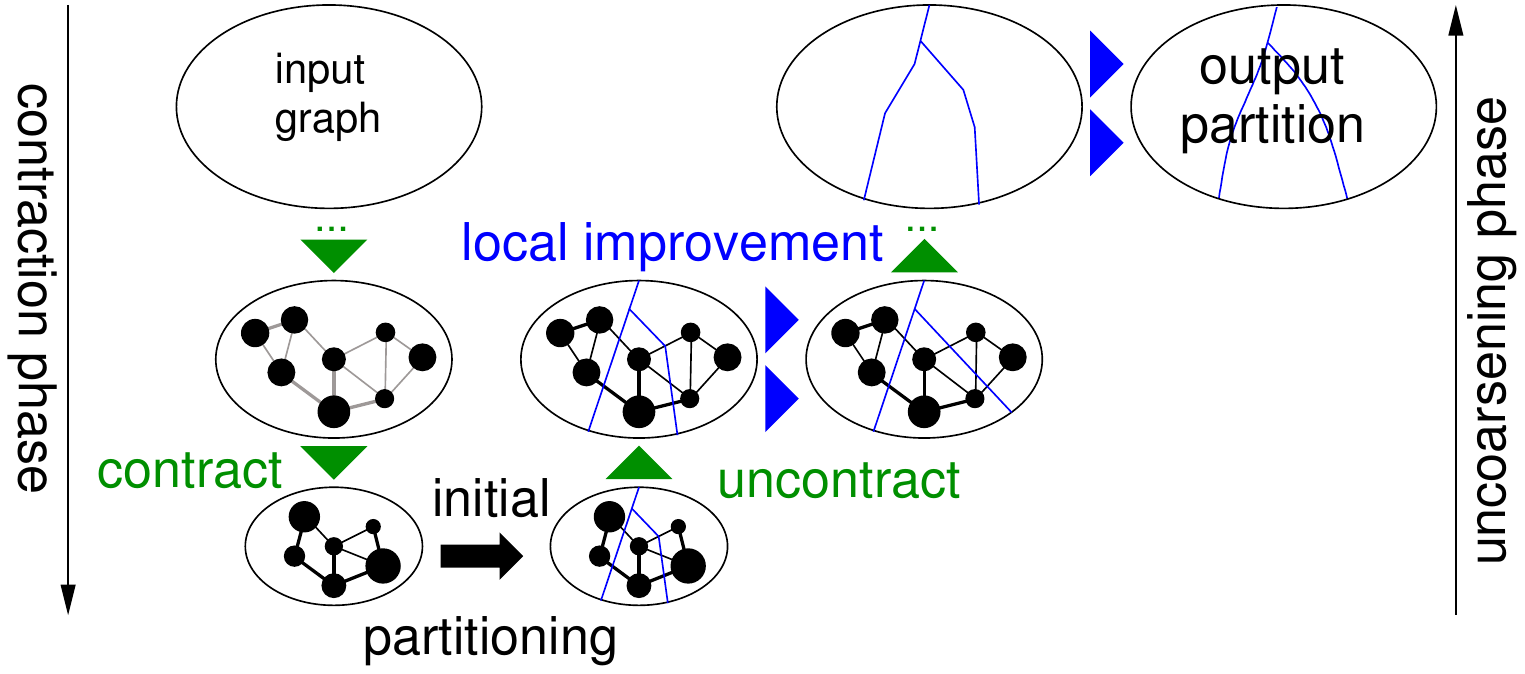}
\caption{Multilevel graph partitioning. The graph is recursively contracted to achieve smaller graphs. After the coarsest graph is initially partitioned, a local search method is used on each level to improve the partitioning induced by the coarser level.}
\label{fig:mgp}
\end{figure}

A successful heuristic for partitioning large graphs is the \emph{multilevel graph partitioning} (MGP) approach depicted in Figure~\ref{fig:mgp},
where the graph is recursively \emph{contracted} to achieve smaller graphs which should reflect the same basic structure as the input graph. After applying an \emph{initial partitioning} algorithm to the smallest graph, the contraction is undone and, at each level, a
\emph{local search} method is used to improve the partitioning induced by the coarser level. 

The main contributions of this paper are a scalable parallelization of the size-constrained label propagation algorithm and an integration into a multilevel framework that enables us to partition large complex networks. 
The parallel size-constrained label propagation algorithm is used to compute a graph clustering which is contracted. 
This is repeated until the graph is small enough. 
The coarsest graph is then partitioned by the coarse-grained distributed evolutionary algorithm KaFFPaE~\cite{kaffpaE}.
During uncoarsening the size-constraint label propagation algorithm is used as a simple, yet effective, parallel local search algorithm. 

The presented scheme speeds up computations and improves solution quality on graphs that have a very irregular structure such as social networks or web graphs.
For example, a variant of our algorithm is able to compute a partition of a web graph with billions of edges in only a few seconds while producing high quality solutions. 

We organize the paper as follows.
We begin in Section~\ref{s:preliminaries} by introducing basic concepts and outlining related work. 
Section~\ref{s:sequentialcontraction} reviews the recently proposed cluster contraction algorithm \cite{pcomplexnetworksviacluster} to partition complex networks, which is parallelized in this work.
The main part of the paper is Section~\ref{s:parallelization}, which covers the parallelization of the size-constrained label propagation algorithm, the parallel contraction and uncontraction algorithm, as well as the overall parallel system.
A summary of extensive experiments to evaluate the algorithm's performance is presented in Section~\ref{s:experiments}. 
Finally, we conclude in Section~\ref{s:conclusion}.

\vfill
\pagebreak
\section{Preliminaries}\label{s:preliminaries}
\subsection{Basic concepts}
Let  $G=(V=\{0,\ldots, n-1\},E,c,\omega)$ be an undirected graph 
with edge weights $\omega: E \to \MdR_{>0}$, node weights
$c: V \to \MdR_{\geq 0}$, $n = |V|$, and $m = |E|$.
We extend $c$ and $\omega$ to sets, i.e.,
$c(V')\Is \sum_{v\in V'}c(v)$ and $\omega(E')\Is \sum_{e\in E'}\omega(e)$.
$\Gamma(v)\Is \setGilt{u}{\set{v,u}\in E}$ denotes the neighbors of $v$.
A node $v \in V_i$ that has a neighbor $w \in V_j, i\neq j$, is a \emph{boundary node}. 
We are looking for \emph{blocks} of nodes $V_1$,\ldots,$V_k$ 
that partition $V$, i.e., $V_1\cup\cdots\cup V_k=V$ and $V_i\cap V_j=\emptyset$
for $i\neq j$. The \emph{balancing constraint} demands that 
$\forall i\in \{1..k\}\gilt c(V_i) \leq L_{\max} := (1+\epsilon)\lceil\frac{c(V)}{k}\rceil$
for some imbalance parameter $\epsilon$. 
The objective is to minimize the total \emph{cut} $\sum_{i<j}w(E_{ij})$ where 
$E_{ij}\Is\setGilt{\set{u,v}\in E}{u\in V_i,v\in V_j}$. 
We say that a block $V_i$ is \emph{underloaded} if $|V_i| < L_{\max}$ and \emph{overloaded} if $|V_i| > L_{\max}$. 

A clustering is also a partition of the nodes, however, $k$ is usually not given in advance and the balance constraint is removed. 
A size-constrained clustering constrains the size of the blocks of a clustering by a given upper bound $U$ such that $c(V_i) \leq U$. 
Note that by adjusting the upper bound one can somehow control the number of blocks of a feasible clustering. 
For example, when using $U=1$, the only feasible size-constrained clustering in an unweighted graph is the clustering where each node forms a block on its own.  

An abstract view of the partitioned graph is the so-called \emph{quotient graph}, in which nodes represent blocks and edges are induced by connectivity between blocks. 
The \emph{weighted} version of the quotient graph has node weights which are set to the weight of the corresponding block and edge weights which are equal to the weight of the edges that run between the respective blocks. 

By default, our initial inputs will have unit edge and node weights. 
However, even those will be translated into weighted problems in the course of the multilevel algorithm.
In order to avoid tedious notation, $G$ will denote the current state of the graph before and after a (un)contraction in the multilevel scheme throughout this paper.
\subsection{Related Work}
\label{s:related}
There has been a \emph{huge} amount of research on graph partitioning so that we refer the reader to \cite{schloegel2000gph,GPOverviewBook,SPPGPOverviewPaper} for most of the material. 
Here, we focus on issues closely related to our main contributions. 
All general-purpose methods that are able to obtain good partitions for large real-world graphs are based on the multilevel principle. 
The basic idea can be traced back to multigrid
solvers for solving systems of linear equations \cite{Sou35} but
more recent practical methods are based on mostly graph theoretic aspects, in
particular edge contraction and local search.  
There are many ways to create graph hierarchies such as matching-based schemes \cite{Chaco,Walshaw07,karypis1998fast,Monien2000,Scotch} or variations thereof \cite{Karypis06} and techniques similar to algebraic multigrid \cite{meyerhenke2006accelerating,ChevalierS09,SafroSS12}. We refer the interested reader to the respective papers for more details.
Well-known software packages based on this approach include Chaco \cite{Chaco}, Jostle~\cite{Walshaw07}, Metis \cite{karypis1998fast}, Party~\cite{Monien2000} and Scotch \cite{Scotch}.  
While Chaco and Party are no longer developed and have no parallel version,
the others have been parallelized, too. 

Most probably the fastest available parallel code is the parallel version of Metis, ParMetis \cite{karypis1996parallel}.
The parallel version of Jostle \cite{Walshaw07} applies local search to pairs of neighboring partitions and is restricted to the case where the number of blocks equals the number of processors. 
This parallelization has problems maintaining the balance of the partitions since at
any particular time, it is difficult to say how many nodes are assigned to a
particular block. 
PT-Scotch \cite{ptscotch}, the parallel version of Scotch, is based on
recursive bipartitioning. This is more difficult to parallelize than direct
$k$-partitioning since in the initial bipartition, there is less parallelism
available.  The unused processor power is used by performing several independent
attempts in parallel. The involved communication effort is reduced by considering only nodes
close to the boundary of the current partitioning (band-refinement). 
KaPPa \cite{kappa} is a parallel matching-based MGP algorithm which is also restricted to the case where the number of blocks equals the number of processors used. 
PDiBaP \cite{Meyerhenke12shape} is a multilevel diffusion-based algorithm that is targeted at small to medium scale parallelism with dozens of processors.

As reported by~\cite{tian2013think}, most large-scale graph processing toolkits based on cloud computing use ParMetis or rather 
straightforward partitioning
strategies such as hash-based partitioning. While hashing often leads to acceptable balance, the edge cut obtained for complex
networks is very high. To address this problem, Tian \etal \cite{tian2013think} have recently proposed a partitioning algorithm
for their toolkit Giraph++. The algorithm uses matching-based coarsening and ParMetis on the coarsest graph. This strategy
leads to better cut values than hashing-based
schemes. However, significant imbalance is introduced by their method, so that their results are incomparable to ours.

The label propagation clustering algorithm was initially proposed by Raghavan \etal \cite{labelpropagationclustering}. 
Moreover, the label propagation algorithm has been used to partition networks by Uganer and Backstrom \cite{UganderB13}. The authors do not use a multilevel scheme and rely on a given or random partition  which are improved by combining the unconstrained label propagation approach with linear programming. Hence, the approach does not yield high quality partitionings.
Another distributed algorithm for balanced graph partitioning has been proposed by Rahimian \etal \cite{jabeja}.  
The authors use random initializations as starting point for local search which is basically node swapping.
However, if the initialization is not balanced, the final partition computed by the algorithm will also be imbalanced and the largest graph under consideration has less than 70K nodes.  

Recent work by Kirmani and Raghavan \cite{KirmaniR13} solves a relaxed version of the graph partitioning problem where no strict balance constraint is enforced. The  
 blocks only have to have approximately the same size. Thus the problem is easier than the version of the problem where a strict balance constraint has to be fullfilled. Their approach attempts to obtain information on the graph structure by computing an embedding into the coordinate space using a multilevel graph drawing algorithm. Afterwards partitions are computed using a geometric scheme. 

\vfill
\pagebreak
\subsection{KaHIP}
\label{s:kaHIP}
Within this work, we use the open source multilevel graph partitioning framework KaHIP~\cite{kaHIPHomePage,kabapeE} (Karlsruhe High Quality Partitioning).
More precisely, we employ the distributed evolutionary algorithm KaFFPaE contained therein to create high quality partitions of complex networks at the coarsest level of the hierarchy. 
Hence, we shortly outline the main components of KaHIP. 

KaHIP implements many different algorithms, for example flow-based methods and more-localized local searches within a multilevel framework called KaFFPa, as well as several coarse-grained parallel and sequential meta-heuristics. 
The algorithms in KaHIP have been able to improve the best known partitioning results in the Walshaw Benchmark~\cite{soper2004combined} for many inputs using a short amount of time to create the partitions.
Recently, also specialized methods to partition social networks and web graphs have been included into the framework \cite{pcomplexnetworksviacluster}. In this work, we parallelize the main techniques presented therein which are reviewed in Section~\ref{s:sequentialcontraction}.

\subsubsection*{KaFFPaE}
We now outline details of the evolutionary algorithm, KaFFPaE \cite{kaffpaE}, since we use this algorithm to obtain a partition of the coarsest graph of the hierarchy. 
KaFFPaE is a coarse-grained evolutionary algorithm, \ie each processing element has its own population (set of partitions) and a copy of the graph.
After initially creating the local population, each processor performs combine and mutation operations on the local population/partitions.
The algorithm contains a general combine operator framework provided by modifications of the multilevel framework KaFFPa.
All combine operators can assure that the offspring has a solution quality at least as good as the better of both parents.   

The basic combine operation works as follows:
Let $\mathcal{P}_1$ and $\mathcal{P}_2$ be two partitions of the graph $G$. 
The partitions are both used as input to the multilevel graph partitioner KaFFPa in the following sense. 
First of all, all edges that are cut edges in any of the two input partitions, \ie edges that run between two blocks, are not eligible for contraction during the coarsening phase. 
This means that they are not contracted during the coarsening phase. 

As soon as the coarsening phase is stopped, the better partition is applied to the coarsest graph and used as 
initial partitioning. 
This is possible since we did not contract any cut edge of $\mathcal{P}_1$ or $\mathcal{P}_2$. 
Since local search algorithms guarantee no worsening of the input partition and random tie breaking is used, it is assured that partition quality is not decreased during uncoarsening.
Note that the local search algorithms can effectively exchange good parts of the solution on the coarse levels by moving only a few nodes.  

To exchange individuals between the processors over time, the algorithm is equipped with a scalable communication protocol similar to randomized rumor spreading.
That means that from time to time, the best local partition is sent to a random selection of other processors.
For more details, we refer the reader to \cite{kaffpaE}.

\section{Cluster Contraction}
\label{s:sequentialcontraction}
We now review the basic idea \cite{pcomplexnetworksviacluster} which we chose to parallelize. 
The approach for creating graph hierarchies is targeted at complex network such as social networks and web graphs.
We start by explaining the size-constrained label propagation algorithm, which is used to compute clusterings of the graph.
To compute a graph hierarchy, the clustering is contracted by replacing each cluster by a single node, and the process is repeated recursively until the graph is small.
Due to the way the contraction is defined, it is ensured that a partition of a coarse graph corresponds to a partition of the input network with the same objective and balance. 
Note that cluster contraction is an aggressive coarsening strategy. In contrast to most previous approaches, it can drastically shrink the size of irregular networks.
The intuition behind this technique is that a clustering of the graph (one hopes) contains many edges running inside the clusters and only a few edges running between clusters, which is favorable for the edge cut objective.
Regarding complexity, experiments in \cite{pcomplexnetworksviacluster} indicate already one contraction step can shrink the graph size by orders of magnitude and that the average degree of the contracted graph is smaller than the average degree of the input network. 
Moreover, the clustering algorithm is fast and essentially runs in linear time. 
On the other hand, the clustering algorithm is parallelizable and, by using a different size-constraint, the label propagation algorithm can also be used as a simple strategy to improve a solution on the current level. 
\subsection{Label Propagation with Size Constraints}
\begin{figure}[b]
\begin{center}                                                      
                 \includegraphics[width=.45\textwidth]{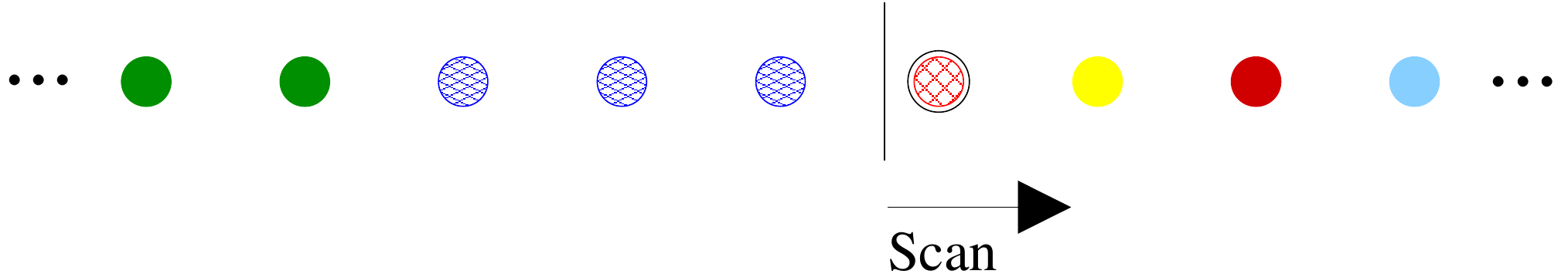}
     \end{center}
     \begin{center}
     \begin{tabular}{ccc}
             \includegraphics[width=2.5cm]{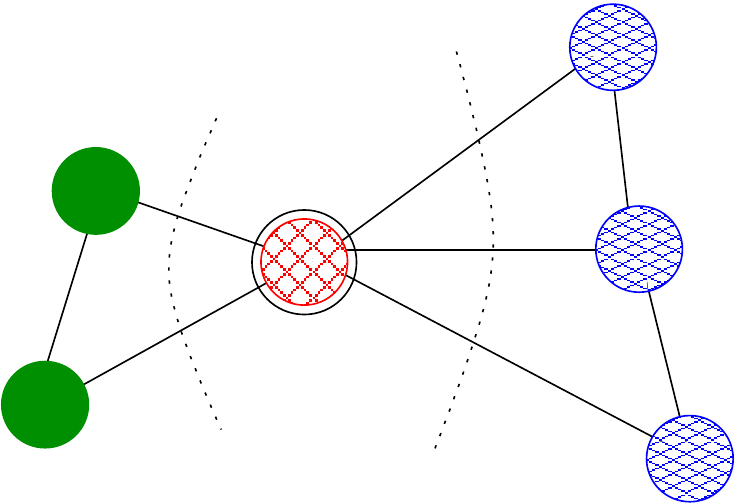} & \begin{minipage}{.025\textwidth}\vspace*{-2cm}\textbf{$\rightarrow$}\end{minipage} &
                \includegraphics[width=2.5cm]{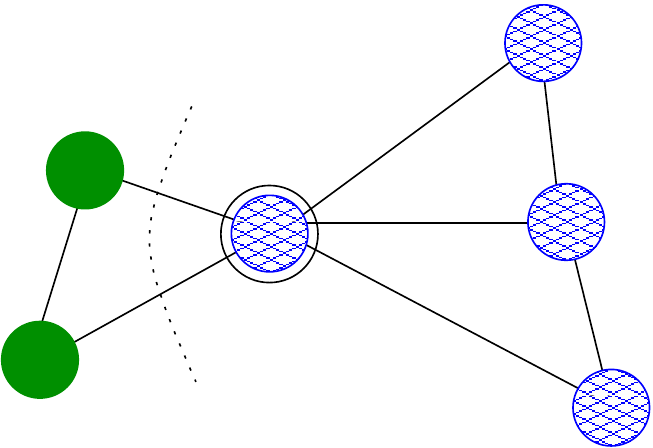}
        \end{tabular}
        \end{center}
        \caption{An example round of the label propagation graph clustering algorithm. Initially each node is in its own block. The algorithm scans all vertices in a random order and moves a node to the block with the strongest connection in its neighborhood.}
\end{figure}

Originally, the \emph{label propagation clustering}  algorithm was proposed by Raghavan \etal \cite{labelpropagationclustering} for graph clustering. 
It is a very fast, near linear-time algorithm that locally optimizes the number of edges cut. We outline the algorithm briefly.  
Initially, each node is in its own cluster/block, \ie the initial block ID of a node is set to its node ID.
The algorithm then works in rounds. 
In each round, the nodes of the graph are traversed in a random order. 
When a node $v$ is visited, it is \emph{moved} to the block that has the strongest connection to $v$, \ie it is moved to the cluster $V_i$ that maximizes $\omega(\{(v, u) \mid u \in N(v) \cap V_i \})$. 
Ties are broken randomly. 
The process is repeated until the process has converged. 
Here, at most $\ell$ iterations of the algorithm are performed, where $\ell$ is a tuning parameter.
One round  of the algorithm can be implemented to run in $\Oh{n+m}$ time. 

\begin{figure}[t]
\centering
\includegraphics[width=.35\textwidth]{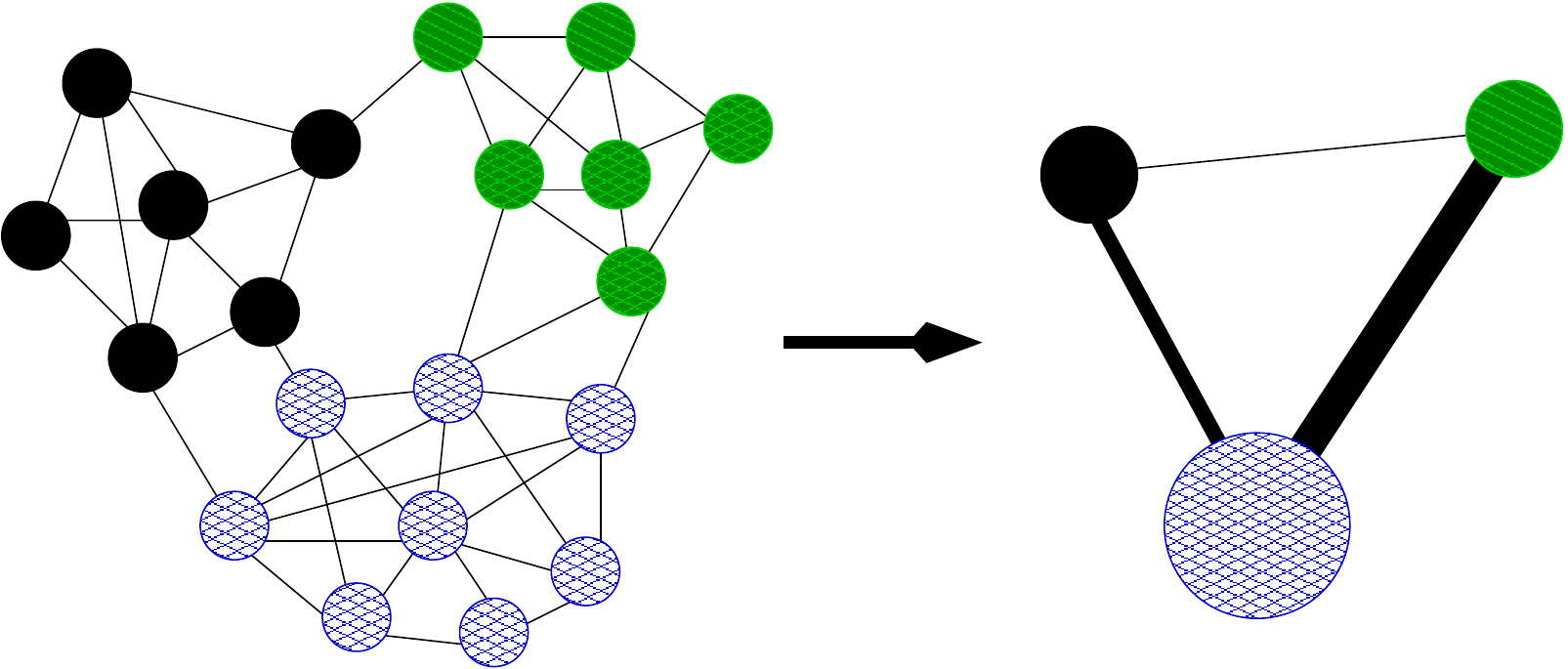}
\caption{Contraction of clusterings. Each cluster of the graph on the left hand side corresponds to a node in the graph on the right hand side. Weights of the nodes and the edges are choosen such that a partition of the coarse graph induces a partition of the fine graph having the same cut and balance.}
\label{fig:clustercontraction}
\end{figure}

The computed clustering is contracted to obtain a coarser graph. 
\emph{Contracting a clustering} works as follows: 
each block of the clustering is contracted into a single node. 
The weight of the node is set to the sum of the weight of all nodes in the original block. 
There is an edge between two nodes $u$ and $v$ in the contracted graph if the
two corresponding blocks in the clustering are adjacent to each other in $G$,
\ie block $u$ and block $v$ are connected by at least one edge.
The weight of an edge $(A,B)$ is set to the sum of the weight of edges that run between block $A$ and block $B$ of the clustering. 
Due to the way contraction is defined, a partition of the coarse graph corresponds to a partition of the finer graph with the same cut and balance. 
An example contraction is shown in Figure~\ref{fig:clustercontraction}. 

In contrast to the original label propagation algorithm~\cite{labelpropagationclustering}, Meyerhenke \etal \cite{pcomplexnetworksviacluster} ensure that each block of the cluster fulfills a size constraint. There are two reason for this.
First, consider a clustering of the graph in which the weight of a block would exceed $(1+\epsilon) \lceil \frac{|V|}{k} \rceil$. 
After contracting this clustering, it would be impossible to find a partition of the contracted graph that fulfills the balance constraint.
Secondly, it has been shown that using more balanced graph hierarchies is beneficial when computing high quality graph partitions~\cite{kappa}.
To ensure that blocks of the clustering do not become too large, an upper bound $U := \max( \max_v c(v), W)$ on the size of the blocks is introduced, where $W$ is a parameter that will be chosen later. When the algorithm starts to compute a
  graph clustering on the input graph, the constraint is fulfilled since each of the blocks contains exactly one node. 
A neighboring block $V_\ell$ of a node $v$ is called \emph{eligible} if $V_\ell$ will not be overloaded once $v$ is moved to $V_\ell$.
Now when a node $v$ is visited, it is moved to the \emph{eligible block} that has the strongest connection to $v$. 
Hence, after moving a node, the size of each block is still smaller than or equal to $U$.
Moreover, after contracting the clustering, the weight of each node is smaller or equal to $U$.
One round of the modified version of the algorithm can still run in linear time by using an array of size $|V|$ to store the block sizes. Note that when parallelizing the algorithm this is something that needs to be adjusted since storing an array of size $|V|$ on a single processor would cost too much memory.
The parameter $W$ is set to $\frac{L_{\text{max}}}{f}$, where $f$ is a tuning parameter. 
Note that the constraint is rather soft during coarsening, \ie in practice it does no harm if a cluster contains slightly more nodes than the upper bound. We go into more detail in the next section. 

The process of computing a size-constrained clustering and contracting it is repeated recursively. 
As soon as the graph is small enough, it is initially partitioned.
That means  each node of the coarsest graph is assigned to a block. 
Afterwards, the solution is transferred to the next finer level. 
To do this,  a node of the finer graph is assigned to the block of its coarse representative. 
Local improvement methods of KaHIP then try to improve the solution on the current level, \ie reducing the number of edges cut.  

Recall that the label propagation algorithm traverses the nodes in a random order and moves a node to a cluster with the strongest connection in its neighborhood to compute a clustering. 
Meyerhenke \etal \cite{pcomplexnetworksviacluster} have shown that using the ordering induced by the node degree (increasing order) improves the overall solution quality \emph{and} running time.
Using this node ordering means that in the first round of the label propagation algorithm, nodes with small node degree can change their cluster before nodes with a large node degree. 
Intuitively, this ensures that there is already a meaningful cluster structure when the label propagation algorithm chooses the cluster of a high degree node. 
Hence, the algorithm is likely to compute better clusterings of the graph by using node orderings based on node degree. 

By using a different size-constraint -- the constraint $W := L_{\text{max}}$ of the original partitioning problem  -- the label propagation is also used as a simple and fast local search algorithm to improve a solution on the current level \cite{pcomplexnetworksviacluster}. 
However, small modifications to handle overloaded blocks have to be made. 
The block selection rule is modified when the algorithm is used as a local search algorithm in case that 
the current node $v$ under consideration is from an overloaded block $V_\ell$.
In this case it is \emph{moved} to the eligible block that has the strongest connection to $v$ without considering the block $V_\ell$ that it is contained in.
This way it is ensured that the move improves the balance of the partition (at the cost of the number of edges cut).
\section{Parallelization}
\label{s:parallelization}
We now present the main contributions of the paper.
We begin with the distributed memory parallelization of the size-constrained label propagation algorithm and continue with the parallel contraction and uncoarsening algorithm. 
At the end of this section, we describe the overall parallel system. 
\subsection{Parallel Label Propagation}
We shortly outline our parallel graph data structure and the implementation of the methods that handle communication. 
First of all, each processing element (PE) gets a subgraph, \ie a contiguous range of nodes $a..b$, of the whole graph as its input, such that the subgraphs combined correspond to the input graph.
Each subgraph consists of the nodes with IDs from the interval $I:=a..b$ and the edges incident to the nodes of those blocks, as well as the end points of edges which are not in the interval $I$ (so-called ghost or halo nodes). 
This implies that each PE may have edges that connect it to another PE and the number of edges assigned to the PEs might vary significantly.
The subgraphs are stored using a standard adjacency array representation, \ie we have one array to store edges and one array for nodes storing head pointers to the edge array. 
However, the node array is divided into two parts. 
The first part stores local nodes and the second part stores ghost nodes. The method used to keep local node IDs and ghost node IDs consistent is explained in the next paragraph.
Additionally, we store information about the nodes, \ie its current block and its weight.  

Instead of using the node IDs provided by the input graph (called global IDs), each PE maps those IDs to the range $0\, .. \, n_p-1$, where $n_p$ is the number of distinct nodes of the subgraph. Note that this number includes the number of ghost nodes the PE has.
Each global ID $i \in a \, .. \, b$ is mapped to a local node ID $i-a$. The IDs of the ghost nodes are mapped to the remaining $n_p - (b-a)$ local IDs in the order in which they appeared during the construction of the graph structure. 
Transforming a local node ID to a global ID or vice versa, can be done by adding or subtracting $a$. 
We store the global ID of the ghost nodes in an extra array and use a hash table to transform global IDs of ghost nodes to their corresponding local IDs.
Additionally, we store for each ghost node the ID of the corresponding PE, using an array for $\mathcal{O}(1)$ lookups. 

To parallelize the label propagation algorithm, each PE performs the algorithm on its part of the graph. 
Recall, when we visit a node $v$, it is moved to the block that has the strongest connection.
Note that the cluster IDs of a node can be arbitrarily distributed in the range $0\, .. \, n-1$ so that we use a hash map to identify the cluster with the strongest connection.
Since we know that the number of distinct neighboring cluster IDs is bounded by the maximum degree in the graph, we use hashing with linear probing.
At this particular point of the algorithm, it turns out that hashing with linear probing is much faster than using the hash map of the STL. 

During the course of the algorithm,  local nodes can change their block and hence the blocks in which ghost nodes are contained can change as well. 
Since communication is expensive, we do not want to perform communication each time a node changes its block. 
We use the following scheme to \emph{overlap} communication and computation. 
The scheme is organized in phases. 
We call a node \emph{interface node} if it is adjacent to at least one ghost node. The PE associated with the ghost node is called adjacent PE.
Each PE stores a separate send buffer for all adjacent PEs.
During each phase, we store the block ID of interface nodes that have changed into the send buffer of each adjacent PE of this node. 
Communication is then implemented asynchronously. 
In phase $\kappa$, we send the current updates to our adjacent PEs and receive the updates of the adjacent PEs from round $\kappa-1$, for $\kappa>1$. 
Note that in case the label propagation algorithm has converged, \ie no node changes its block any more, the communication volume is really small. 

The degree-based node ordering approach of the label propagation algorithm that is used during coarsening is parallelized by considering only the local nodes for this ordering. 
In other words, the ordering in which the nodes are traversed on a PE is determined by the node degrees of the local nodes of this PE. During uncoarsening random node ordering is used.
\subsection{Balance/Size Constraint}
\label{ss:balanceconstraint}
Recall that we use the size-constrained label propagation algorithm during coarsening using $\frac{L_{\max}}{f}$ as a size constraint and during uncoarsening using $L_{\max}$ as a size constraint.
Maintaining the balance of blocks is somewhat more difficult in the parallel case than in the sequential case.
We use two different approaches to maintain balance, one of which is used during coarsening and the other one is used during uncoarsening.
The reason for this is that during coarsening there is a large number of blocks and the constraint is rather soft, whereas during uncoarsening the number of blocks is small and the constraint is tight.

We maintain the balance of different blocks \emph{during coarsening} as follows. 
Roughly speaking, a PE maintains and updates only the local amount of node weight of the blocks of its local and ghost nodes.
Due to the way the label propagation algorithm is initialized, each PE knows the exact weights of the blocks of local nodes and ghost nodes in the beginning. The label propagation then uses the local information to bound the block weights. Once a node changes its block, the local block weight is updated.
Note that this does not involve additional amounts of communication.
We decided to use this localized approach since the balance constraint is not tight during coarsening.
More precisely, the bound on the cluster sizes during coarsening is a tuning parameter and the overall performance of the system does not depend on the exact choice of the parameter.

\emph{During uncoarsening} we use a different approach since the number of blocks is much smaller and it is unlikely that the previous approach yields a feasible partition in the end. 
This approach is similar to the approach that is used within ParMetis~\cite{karypis1996parallel}.
Initially, the exact block weights of all $k$ blocks are computed locally.
The local block weights are then aggregated and  broadcast to all PEs. Both can be done using one allreduce operation.
Now each PE knows the global block weights of all $k$ blocks.
The label propagation algorithm then uses this information and locally updates the weights.
For each block, a PE maintains and updates the total amount of node weight that local nodes contribute to the block weights. 
Using this information, one can restore the exact block weights with one allreduce operation which is done at the end of each computation phase. 
Note that this approach would not be feasible during coarsening since there are $n$ blocks in the beginning of the algorithm and each PE holds the block weights of all blocks. 
\subsection{Parallel Contraction and Uncoarsening}
The \emph{parallel contraction} algorithm works as follows. 
After the parallel size-constrained label propagation algorithm has been performed, each node is assigned to a cluster. 
Recall the definition of our general contraction scheme.
Each of the clusters of the graph corresponds to a coarse node in the coarse graph and the weight of this node is set to the total weight of the nodes that are in that cluster. Moreover, there is an edge between two coarse nodes iff there is an edge between the respective clusters and the weight of this edge is set to the total weight of the edges that run between these clusters in the original graph. 

In the parallel scheme, the IDs of the clusters on a PE can be arbitrarily distributed in the interval $0\, ..\, n-1$, where $n$ is the total number of nodes of the input graph of the current level. 
Consequently, we start the parallel contraction algorithm by finding the number of distinct cluster IDs which is also the number of coarse nodes. 
To do so, a PE $p$ is assigned to count the number of distinct cluster IDs in the interval $I_p:= p\lceil \frac{n}{P} \rceil+1 \, .. \, (p+1)\lceil \frac{n}{P} \rceil$, where $P$ is the total number of PEs used. 
That means each PE $p$ iterates over its local nodes, collects cluster IDs $a$ that are not local, \ie $a \not \in I_p$, and then sends the non-local cluster IDs to the responsible PEs. 
Afterwards, a PE counts the number of distinct local cluster IDs so that the number of global distinct cluster IDs can be derived easily by using a reduce operation.

Let $n'$ be the global number of distinct cluster IDs. 
Recall that this is also the number of coarse nodes after the contraction has been performed. 
The next step in the parallel contraction algorithm is to compute a mapping $q: 0 \, .. 
\,  n-1 \to 0 \, .. \, n'-1$ which maps the current cluster IDs to a contiguous interval over all PEs. This mapping can be easily computed in parallel by computing a prefix sum over the number of distinct local cluster IDs a PE has.
Once this is done, we compute the mapping $C: 0 \, .. \, n-1 \to 0 \, .. \, n'-1$ which maps a node ID of $G$ to its coarse representative. 
Note that, if a node $v$ is in cluster $V_\ell$ after the label propagation algorithm has converged, then $C(v) = q(\ell)$. 
After computing this information locally, we also propagate the necessary parts of the mapping to neighboring PEs so that we also know the coarse representative of each ghost node.
When the contraction algorithm is fully completed, PE $p$ will be \emph{responsible} for the subgraph $p\lceil \frac{n'}{P} \rceil+1\, .. \, (p+1)\lceil \frac{n'}{P} \rceil$ of the coarse graph.
To construct the final coarse graph, we first construct the weighted quotient graph of the local subgraph of $G$ using hashing. 
Afterwards, each PE sends an edge $(u,v)$ of the local quotient graph, including its weight and the weight of its source node, to the responsible PE. 
After all edges are received, a PE can construct its coarse subgraph locally. 

The implementation of the \emph{parallel uncoarsening} algorithm is simple. Each PE knows the coarse node for all its nodes in its subgraph (through the mapping $C$). Hence, a PE requests the block ID of a coarse representative of a fine node from the PE that holds the respective coarse node.
\subsection{Miscellanea}

\subsubsection*{Iterated Multilevel Schemes}
A common approach to obtain high quality partitions is to use a multilevel algorithm multiple times using different random seeds 
and use the best partition that has been found.
However, one can do better by transferring the solution of the previous multilevel iteration down the hierarchy. 
In the graph partitioning context, the notion of V-cycles was introduced by Walshaw \cite{walshaw2004multilevel}. More recent
work augmented them to more complex cycles~\cite{kaffpa}. 
These previous works use matching-based coarsening with cut edges not being  matched (and hence cut edges are not contracted).
Thus, a given partition on the finest level can be used as initial partition of the coarsest graph (having the same balance and cut as the partition of the finest graph).   

Iterated V-cycles are also used within clustering-based coarsening by Meyerhenke \etal \cite{pcomplexnetworksviacluster}.
To adopt  the iterated multilevel technique for this coarsening scheme, it has to be ensured that cut edges are not contracted after the first multilevel iteration.
This is done by modifying the label propagation algorithm such that each cluster of the computed clustering is a subset of a block of the input partition. 
In other words, each cluster only contains nodes of one unique block of the input partition. 
Hence, when contracting the clustering, every cut edge of the input partition will remain.
Recall that the label propagation algorithm initially puts each node in its own block so that in the beginning of the algorithm each cluster is a subset of one unique block of the input partition. 
This property is kept during the course of the label propagation algorithm by restricting the movements of the label propagation algorithm, \ie we move a node to an eligible cluster with the strongest connection in its neighborhood that is in the same block of the input partition as the node itself. 
We do the same in our parallel approach to realize V-cycles.
\begin{figure}[t]
\includegraphics[width=.475\textwidth]{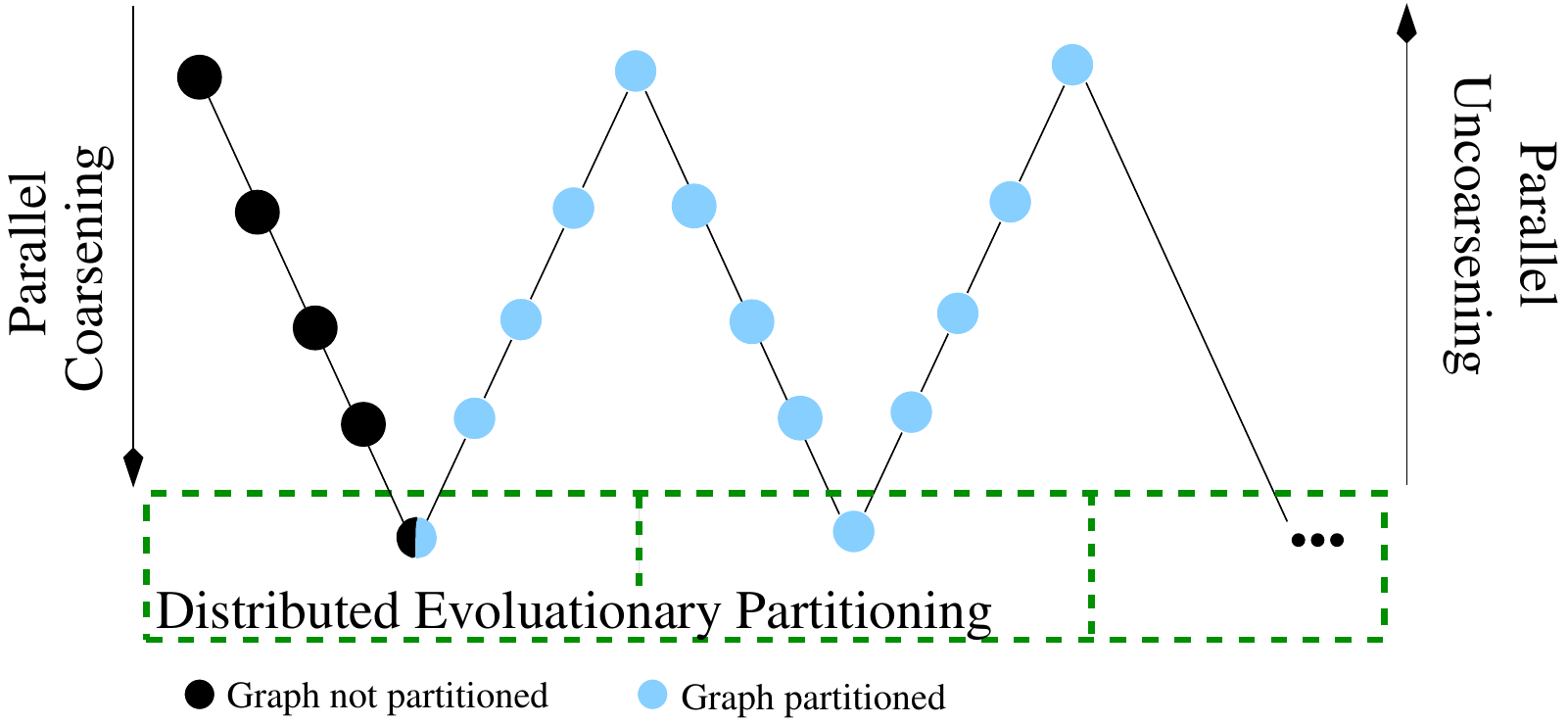}
\caption{The overall parallel system. It uses the parallel cluster coarsening algorithm, the coarse-grained distributed evolutionary algorithm KaFFPaE to partition the coarsest graph  and parallel uncoarsening/local search. After the first iteration of the multilevel scheme the input partition is used as a partition of the coarsest graph and used as a starting point by the evolutionary algorithm.}
\label{fig:}
\end{figure}

\subsection{The Overall Parallel System}
The overall parallel system works as follows.
We use $\ell$ iterations of the parallel size-constrained label propagation algorithm to compute graph clusterings and contract them in parallel. 
We do this recursively until the remaining graph has $10\,000 \cdot k$ nodes left, where $k$ is the number of blocks that the input network should be partitioned in. The distributed coarse graph is then collected on each PE, \ie each PE has a copy of the complete coarsest graph. 
We use this graph as input to the coarse-grained distributed evolutionary algorithm KaFFPaE, to obtain a high quality partition of it. 
KaFFPaE uses modified combine operations that also use the clustering-based coarsening scheme from above.
The best solution of the evolutionary algorithm is then broadcast to all PEs which transfer the solution to their local part of the distributed coarse graph.
Afterwards, we use the parallel uncoarsening algorithm to transfer the solution of the current level to the next finer level and apply $r$ iterations of the parallel label propagation algorithm with the size constraints of the original partitioning problem (setting $W=(1+\epsilon) \lceil\frac{|V|}{k}\rceil$) to improve the solution on the current level. 
We do this recursively on each level and obtain a $k$-partition of the input network in the end. 
If we use iterated V-cycles, we use the given partition of the coarse graph as input to the evolutionary algorithm. 
More precisely,  one individual of the population is the input partition on each PE.
This way it is ensured that the evolutionary algorithm computes a partition that is at least as good as the given partition.

\section{Experiments}
\label{s:experiments}
In this section we carefully evaluate the performance of the proposed algorithm. We start by presenting our methodology, the systems used for the evaluation and the benchmark set that we used. We then look into solution quality as well as weak and strong scalability comparing our algorithm to ParMetis, which is probably the most widely used parallel partitioning algorithm.

\subsection{Methodology}
We have implemented the algorithm described above using C++ and MPI. 
Overall, our parallel program consists of about 7000 lines of code (not including the source of KaHIP 0.61). 
We compiled it using g++ 4.8.2 and OpenMPI 1.6.5.
For the following comparisons we used ParMetis 4.0.3. All programs have been compiled using 64 bit index data types. 
We also ran PT-Scotch 6.0.0, but the results have been consistently worse in terms of solution quality and running time compared to the results computed by ParMetis, so that we do not present detailed data for PT-Scotch. 
Our default value for the allowed imbalance is 3\% since this is one of the values used in \cite{walshaw2000mpm} and the default value in Metis.
By default we perform ten repetitions for each configuration of the algorithm using different random seeds for initialization and report the arithmetic average of computed cut size, running time and the best cut found. 
When further averaging over multiple instances, we use the geometric mean in order to give every instance a comparable influence on the final score.  
Unless otherwise stated, we use the following factor $f$ of the size-constraint (see Section~\ref{ss:balanceconstraint} for the definition): during the first V-cycle the factor $f$ is set to $14$ on social networks as well as web graphs and to $20\,000$ on mesh type networks. In later V-cycles we use a random value  $f \in_{\text{rnd}}[10,25]$ to increase the diversifaction of the algorithms.
Our experiments mainly focus on the case $k=2$ to save running time and to keep the experimental evaluation simple. 
Moreover, we used $k=16$ for the number of blocks when performing the weak scalability experiments in Section~\ref{s:expweakscalability}.

\subsubsection*{Algorithm Configurations} 
Any multilevel algorithm has a considerable number of choices between
algorithmic components and tuning parameters. 
We define two ``good'' choices: the \Id{fast} setting aims at a low execution
time that still gives good partitioning quality and the \Id{eco}
setting targets even better partitioning quality without investing an outrageous amount
of time.  When not otherwise mentioned, we
use the \Id{fast} parameter setting.  

The \Id{fast} configuration of our algorithm uses three label propagation iterations during coarsening and six during refinement. 
We also tried larger amounts of label propagation iterations during coarsening, but did not observe a significant impact on solution quality.
This configuration gives the evolutionary algorithm only enough time to compute the initial population and performs two V-cycles.
The \Id{eco} configuration of our algorithm uses three label propagation iterations during coarsening and six label propagation iterations during refinement as well. 
Time spent during initial partitioning is dependent on the number of processors used. 
To be more precise, when we use one PE, the evolutionary algorithm has $t_1=2^{11}s$ to compute a partition of the coarsest graph during the first V-cycle.
When we use $p$ PEs, then it gets time $t_p=t_1/p$ to compute a partition of an instance. This configuration performs five V-cycles.
There is also a \Id{minimal} variant of the algorithm, which is similar to the \Id{fast} configuration but only performs one V-cycle.
We use this variant of the algorithm only once -- to create a partition of the largest web graph uk-2007 on machine~B (described
below).

\subsubsection*{Systems} 
We use two different systems for our experimental evaluation.
\emph{System A} is mainly used for the evaluation of the solution quality of the different algorithms in Table~\ref{tab:bipartitioningresults}. It is equipped with four Intel Xeon E5-4640 Octa-Core processors (Sandy Bridge) running at a clock speed of 2.4 GHz. The machine has 512 GB main memory, 20 MB L3-Cache  and 8x256 KB L2-Cache. 
\emph{System B} is a cluster where each node is equipped with two Intel Xeon E5-2670 Octa-Core processors (Sandy Bridge) which run at a clock speed of 2.6 GHz. 
Each node has 64 GB local memory, 20 MB L3-Cache and 8x256 KB L2-Cache.
All nodes have local disks and are connected by an InfiniBand 4X QDR interconnect, which is characterized by its very low latency of about 1 microsecond and a point to point bandwidth between two nodes of more than 3700 MB/s. We use machine $B$ for the scalability experiments in Section~\ref{s:expweakscalability}.

\subsubsection*{Instances}
We evaluate our algorithms on graphs collected from \cite{benchmarksfornetworksanalysis,UFsparsematrixcollection,BoVWFI,snap}. 
Table~\ref{tab:scalefreegraphstable} summarizes the main properties of the benchmark set. 
Our benchmark set includes a number of graphs from numeric simulations as well as social networks and web graphs.  
Moreover, we use the two graph families \Id{rgg} and \Id{del} for comparisons. 
\Id{rgg$X$} is a \emph{random geometric graph} with
$2^{X}$ nodes where nodes represent random points in the unit square and edges
connect nodes whose Euclidean distance is below $0.55 \sqrt{ \ln n / n }$.
This threshold was chosen in order to ensure that the graph is almost certainly connected. 
The largest graph of this class is \Id{rgg31}, which has about 21.9 billion edges.
\Id{del$X$} is a Delaunay triangulation of $2^{X}$
random points in the unit square. 
The largest graph of this class is \Id{del31}, which has about 6.4 billion edges.
Most of these graphs are available at the 10th DIMACS Implementation Challenge~\cite{dimacschallengegraphpartandcluster} website. 
The largest graphs (with $2^{26}$ to $2^{31}$ nodes) of these families have been generated using modified code taken from \cite{kappa}. 
We will make these graphs available on request.

\begin{table}[t]
\centering
\caption{Basic properties of the benchmark set with a rough type classification. S stands for social or web graphs, M is used for mesh type networks.}
 \label{tab:scalefreegraphstable}
\begin{tabular}{|l|r|r||r||r|}
\hline
graph & $n$ & $m$ & Type & Ref. \\
\hline
\hline
        \multicolumn{4}{|c|}{Large Graphs} \\
\hline
amazon                                     & $\approx$407K              & $\approx$2.3M        &S& \cite{snap}\\
eu-2005                                    & $\approx$862K              & $\approx$16.1M       &S& \cite{benchmarksfornetworksanalysis}\\
youtube                                    & $\approx$1.1M              & $\approx$2.9M        &S& \cite{snap}\\
in-2004                                    & $\approx$1.3M              & $\approx$13.6M       &S& \cite{benchmarksfornetworksanalysis}\\
packing                                    & $\approx$2.1M              & $\approx$17.4M       &M& \cite{benchmarksfornetworksanalysis}\\
enwiki                                     & $\approx$4.2M              & $\approx$91.9M       &S& \cite{webgraphWS} \\
channel                                    & $\approx$4.8M              & $\approx$42.6M       &M& \cite{benchmarksfornetworksanalysis}\\
hugebubble-10                              & $\approx$18.3M             & $\approx$27.5M       &M& \cite{benchmarksfornetworksanalysis}\\
nlpkkt240                                  & $\approx$27.9M             & $\approx$373M        &M& \cite{UFsparsematrixcollection}\\
uk-2002                                    & $\approx$18.5M             & $\approx$262M        &S& \cite{webgraphWS} \\
del26                                      & $\approx$67.1M             & $\approx$201M        &M& \cite{kappa} \\
rgg26                                      & $\approx$67.1M             & $\approx$575M        &M& \cite{kappa} \\
\hline                                                                                          
\multicolumn{4}{|c|}{Larger Web Graphs} \\                                                      
\hline                                                                                          
arabic-2005                                & $\approx$22.7M             & $\approx$553M        &S& \cite{webgraphWS}  \\
sk-2005                                    & $\approx$50.6M             & $\approx$1.8G        &S& \cite{webgraphWS}    \\
uk-2007                                    & $\approx$105.8M            & $\approx$3.3G        &S& \cite{webgraphWS}  \\
\hline                                                                                          
\multicolumn{4}{|c|}{Graph Families} \\                                                         
\hline                                                                                          
delX                                       & [$2^{19}, \ldots, 2^{31}$] & $\approx$1.5M--6.4G  &M& \cite{kappa}\\
\hline                                                                                          
rggX                                       & [$2^{19}, \ldots, 2^{31}]$ & $\approx$3.3M--21.9G &M& \cite{kappa}\\
\hline
\end{tabular}
\end{table}

\subsection{Main Results and Comparison to ParMetis}
\begin{table*}[htb]
\small
\centering
\caption{Average performance (cut and running time) and best result achieved by different partitioning algorithms. Results are for the bipartitioning case $k=2$. All tools used 32 PEs of machine A. Results indicated by a * mean that the amount of memory needed by the partitioner exceeded the amount of memory available on that machine when 32 PEs are used (512GB RAM). The ParMetis result on arabic has been obtained using 15 PEs (the largest number of PEs so that ParMetis could solve the instance).}
\label{tab:bipartitioningresults}
\begin{tabular}{|l||r|r|r||r|r|r||r|r|r|}
\hline
 algorithm  &  \multicolumn{3}{c||}{ParMetis} & \multicolumn{3}{c||}{Fast} &  \multicolumn{3}{c|}{Eco} \\
\hline
graph &    avg. cut & best cut & $t$[s]& avg. cut & best cut & $t$[s] &  avg. cut & best cut & $t$[s]\\
\hline
\hline
amazon      & \numprint{48104}   & \numprint{47010}   & \numprint{0.49}   & \numprint{46641}   & \numprint{45872}   & \numprint{1.85}   & \numprint{44703}   & \textbf{\numprint{44279}}   & \numprint{71.04}\\
eu-2005     & \numprint{33789}   & \numprint{24336}   & \numprint{30.60}  & \numprint{20898}   & \numprint{18404}   & \numprint{1.63}   & \numprint{18565}   &  \textbf{ \numprint{18347}}   & \numprint{70.04}    \\
youtube     & \numprint{181885}  & \numprint{171857}  & \numprint{6.10}   & \numprint{174911}  & \numprint{171549}  & \numprint{8.74}   & \numprint{167874}  & \textbf{\numprint{164095}}  & \numprint{105.87} \\
in-2004     & \numprint{7016}    & \numprint{5276}    & \numprint{3.43}   & \numprint{3172}    & \numprint{3110}    & \numprint{1.38}   & \numprint{3027}    & \textbf{\numprint{2968}}    & \numprint{69.19}    \\
packing     & \numprint{11991}   & \numprint{11476}   & \numprint{0.24}   & \numprint{10185}   & \numprint{9925}    & \numprint{1.84}   & \numprint{9634}    & \textbf{\numprint{9351}}    & \numprint{68.69}    \\
enwiki      & \numprint{9578551} & \numprint{9553051} & \numprint{326.92} & \numprint{9622745} & \numprint{9565648} & \numprint{157.32} & \numprint{9559782} & \textbf{\numprint{9536520}} & \numprint{264.64}   \\
channel     & \numprint{48798}   & \textbf{\numprint{47776}}   & \numprint{0.55}   & \numprint{56982}   & \numprint{55959}   & \numprint{2.71}   & \numprint{52101}   & \numprint{50210}   & \numprint{71.95}    \\
hugebubbles & \numprint{1922}    & \numprint{1854}    & \numprint{4.66}   & \numprint{1918}    & \numprint{1857}    & \numprint{38.00}  & \numprint{1678}    & \textbf{\numprint{1620}}    & \numprint{216.91}   \\
nlpkkt240   & \numprint{1178988} & \textbf{\numprint{1152935}} & \numprint{15.97}  & \numprint{1241950} & \numprint{1228086} & \numprint{35.06}  & \numprint{1193016} & \numprint{1181214} & \numprint{192.78}   \\
uk-2002     & \numprint{787391}  & \numprint{697767}  & \numprint{128.71} & \numprint{434227}  & \numprint{390182}  & \numprint{19.62}  & \numprint{415120}  & \textbf{\numprint{381464}}  & \numprint{146.77}   \\
del26       & \numprint{18086}   & \numprint{17609}   & \numprint{23.74}  & \numprint{17002}   & \numprint{16703}   & \numprint{165.02} & \numprint{15826}   & \textbf{\numprint{15690}}   & \numprint{697.43}   \\
rgg26       & \numprint{44747}   & \numprint{42739}   & \numprint{8.37}   & \numprint{38371}   & \numprint{37676}   & \numprint{55.91}  & \numprint{34530}   & \textbf{\numprint{34022}}   & \numprint{263.81}   \\
arabic-2005 & *\numprint{1078415}& *\numprint{968871}& *\numprint{1245.57} & \numprint{551778}  & \textbf{\numprint{471141}}  & \numprint{33.45}  & \numprint{511316}  & \numprint{475140}  & \numprint{184.01}   \\
sk-2005     & *                  & *                  & *                 & \numprint{3775369} & \numprint{3204125} & \numprint{471.16} & \numprint{3265412} & \textbf{\numprint{2904521}} & \numprint{1688.63}  \\
uk-2007     & *                  & *                  & *                 & \numprint{1053973} & \numprint{1032000} & \numprint{169.96} & \numprint{1010908} & \textbf{\numprint{981654}}  & \numprint{723.42}   \\
\hline
\end{tabular}

\end{table*}

\label{s:expsolutionquality}
In this section we compare variants of our algorithm against ParMetis in terms of solution quality, running time as well as weak and strong scalability.
We start with the comparison of solution quality (average cut, best cut) and average running time on most of the graphs from Table~\ref{tab:scalefreegraphstable} when 32 PEs of machine A are used. Table~\ref{tab:bipartitioningresults} gives detailed results per instance.

First of all, ParMetis could not solve the largest instances in our benchmark set, arabic, sk-2005 and uk-2007 when 32 PEs of machine A are used. This is due to the fact that ParMetis cannot coarsen the graphs effectively so that 
the coarsening phase is stopped too early. 
Since the smallest graph is replicated on each of the PEs, the amount of memory needed by ParMetis is larger than the amount of memory provided by the machine (512GB RAM). 
For example, when the coarsening phase of ParMetis stops on the instance uk-2007, the coarsest graph still has more than 60M vertices. This is less than a factor of two reduction in graph size compared to the input network (the same holds for the number of edges in the coarse graph). 
The same behaviour is observed on machine B, where even less memory per PE is available. 
Contrarily, our algorithm is able to shrink the graph size significantly. 
For instance, after the first contraction step, the graph is already two orders of magnitude smaller and contains a factor of 300 less edges than the input graph uk-2007.
We also tried to use a smaller amount of PEs for ParMetis. 
It turns out that ParMetis can partition arabic when using 15 PEs cutting nearly twice as many edges and consuming thirty-seven times more running time than our \Id{fast} variant. Moreover, ParMetis could not solve the instance sk-2005 and uk-2007 for any number of PEs. 

When only considering the networks that ParMetis could solve  in Table~\ref{tab:bipartitioningresults}, our \Id{fast} and \Id{eco} configuration compute cuts that are 19.2\% and 27.4\% smaller on average than the cuts computed by ParMetis, respectively.
On average, \Id{fast} and \Id{eco} need more time to compute a partition. 
However, there is a well defined \emph{gap} between mesh type networks that usually do not have a community structure to be found and contracted by our algorithm, and social networks as well as web graphs, which our algorithm targets.
Considering only social networks and web graphs, our \Id{fast} algorithm is more than a factor two faster on average and improves the cuts produced by ParMetis by 38\% (the \Id{eco} configuration computes cuts that are 45\% smaller than the cuts computed by ParMetis).
The largest speedup over ParMetis in Table~\ref{tab:bipartitioningresults} was obtained on eu-2005 where our algorithm is more than eighteen times as fast as ParMetis and cuts 61.6\% less edges on average.
In contrast, on mesh type networks our algorithm does not have the same advantage as on social networks.
For example, our \Id{fast} configuration improves on ParMetis only by 2.9\% while needing more than five times as much running time. 
This is due to the fact that this type of network usually has no community structure so that the graph sizes do not shrink as fast.
Still the \Id{eco} configuration computes 11.8\% smaller cuts than ParMetis.
To obtain a fair comparison on this type of networks, we also compare the best cut found by ParMetis against the average cuts found by our algorithms.
While the best cuts on mesh type networks of ParMetis are comparable to the average results of our \Id{fast} configuration, the \Id{eco} configuration still yields 8.2\% smaller cuts.
When partitioning the instances into 32 blocks, improvements are distributed similarly, \ie they are much larger on social networks than on mesh type networks. Overall, our \Id{fast} and \Id{eco} configuration compute 6.8\% and 16.1\% smaller cuts than ParMetis, respectively. Yet, in this comparison
ParMetis simplifies the problem by relaxing it: On some instances it does not respect the balance 
constraint and computes partitions with up to 6\% imbalance. 
\begin{figure}[h]
\vspace*{-.5cm}
\begin{center}
\hspace*{-.03\textwidth}
\includegraphics[width=.5\textwidth]{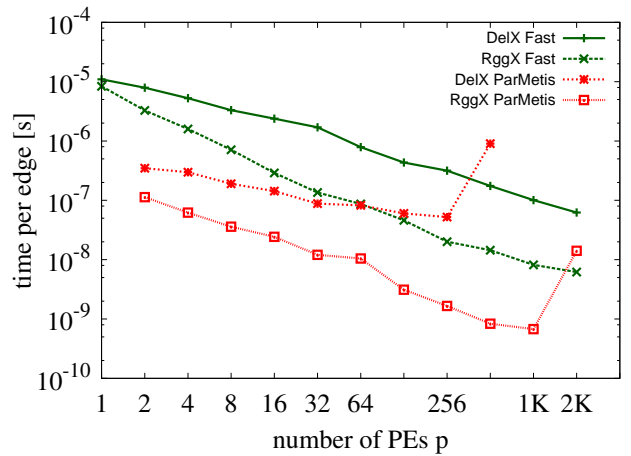}
\end{center}
\vspace*{-.75cm}
\caption{Weak scaling experiments for random geometric graph class \Id{rggX} and the Delaunay triangulation graph class \Id{delX}. When using $p$ PEs, the  instance with $2^{19}p$ nodes from the corresponding graph class was used, \ie when using 2048 cores all algorithms partition the graphs \Id{del30} and \Id{rgg30}. The figure shows the time spend per edge. Sixteen blocks have been used for the partitioning task.}
\label{fig:weakscalingall}
\end{figure}
A table reporting detailed results can be found in the Appendix. 

We now turn to the evaluation of \emph{weak scalability}. These experiments have been performed on the high-performance cluster (machine B).
To evaluate weak scalability, we use the two graph families \Id{rgg$X$} and \Id{del$X$}, and use $k=16$ for the number of blocks for the partitioning task.  
Moreover, we focus on the \Id{fast} configuration of our algorithm and 
ParMetis 
to save running time. We expect that the scalability of the
 \Id{eco} configuration of our algorithm is similar.
When using $p$ PEs, the  instance with $2^{19}p$ nodes from the corresponding graph class is used, \ie when using 2048 cores, all algorithms partition the graphs \Id{del30} and \Id{rgg30}.
Figure~\ref{fig:weakscalingall} reports the running time per edge of the algorithms under consideration.
Our algorithm shows weak scalability \emph{all the way down} to the largest number of cores used while the running time per edge has a somewhat stronger descent compared to ParMetis.
ParMetis has trouble partitioning the largest Delaunay graphs. 
The largest Delaunay graph that ParMetis could partition was \Id{del28} using 512 cores. 
Considering the instances that ParMetis could solve, our \Id{fast} configuration improves solution quality by 19.5\% on
 random geometric graphs and by 11.5\% on Delaunay triangulations on average. 
Since the running time of the \Id{fast} configuration is slower on both graph families, we again compare the best cut results of ParMetis achieved in ten repetitions against our average 
 results to obtain a fair comparison (in this case ParMetis has a slight advantage in terms of running time). 
Doing so, our algorithm still yields an improvement of 16.8\% on the random geometric graphs and an improvement of 9.5\% on the Delaunay triangulations.
Eventually, ParMetis is slower than the \Id{fast} version of our partitioner. On the largest random geometric graph used during this test, we are about a factor of two faster than ParMetis, while improving the results of ParMetis by 9.5\%. In this case our partitioner needs roughly 65 seconds to compute a 16-partition of the graph. In addition, our algorithm is a factor five faster on the largest Delaunay graph that ParMetis could solve and produces a cut that is 9.5\% smaller than the cut produced by ParMetis. 
\begin{figure}
        \centering
        \vspace*{-.5cm}
        \hspace*{-.03\textwidth}
        \vspace*{-.5cm}
        \includegraphics[width=.5\textwidth]{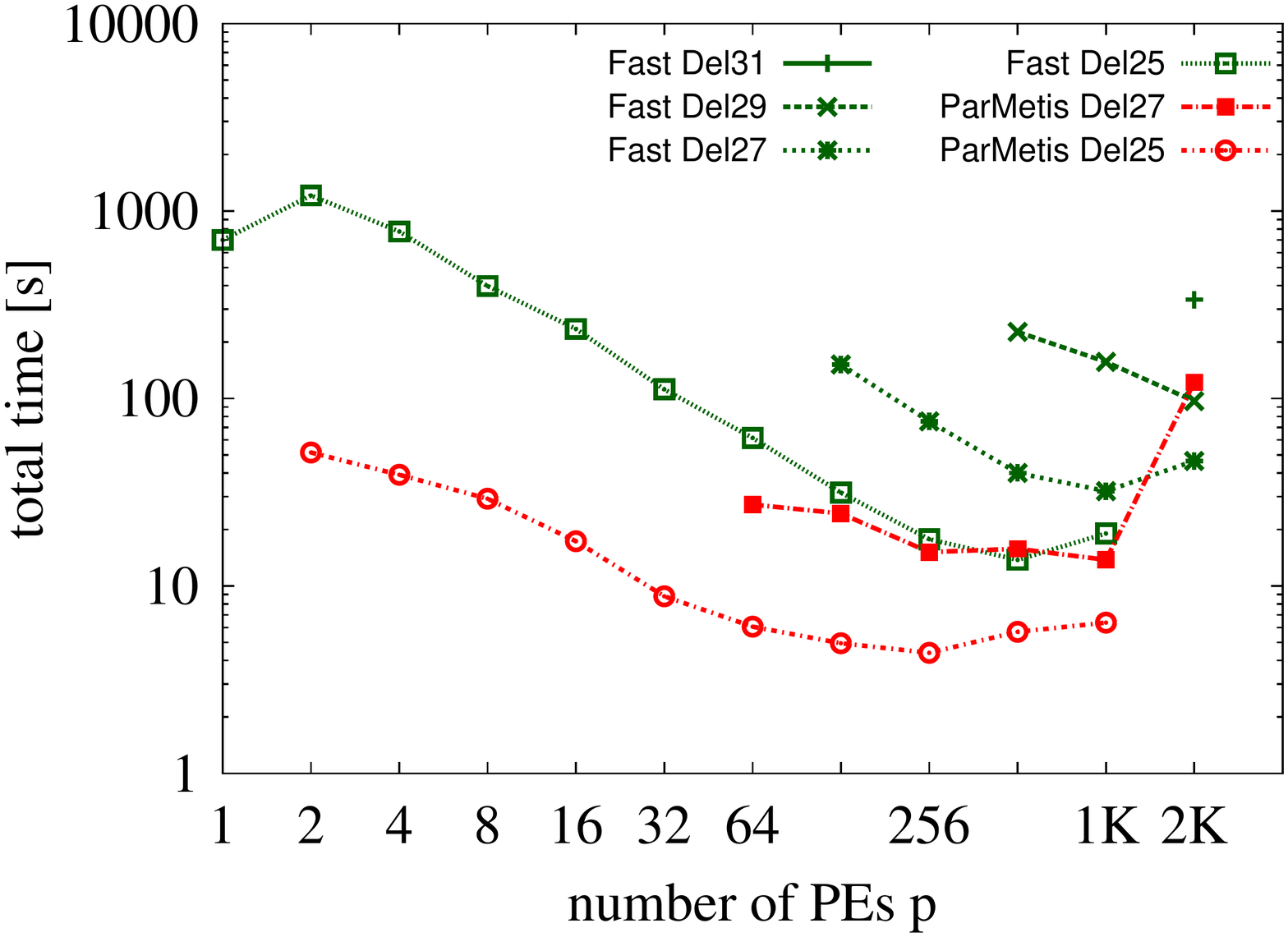}
        \vspace*{-.5cm}
        \hspace*{-.03\textwidth}
        \includegraphics[width=.5\textwidth]{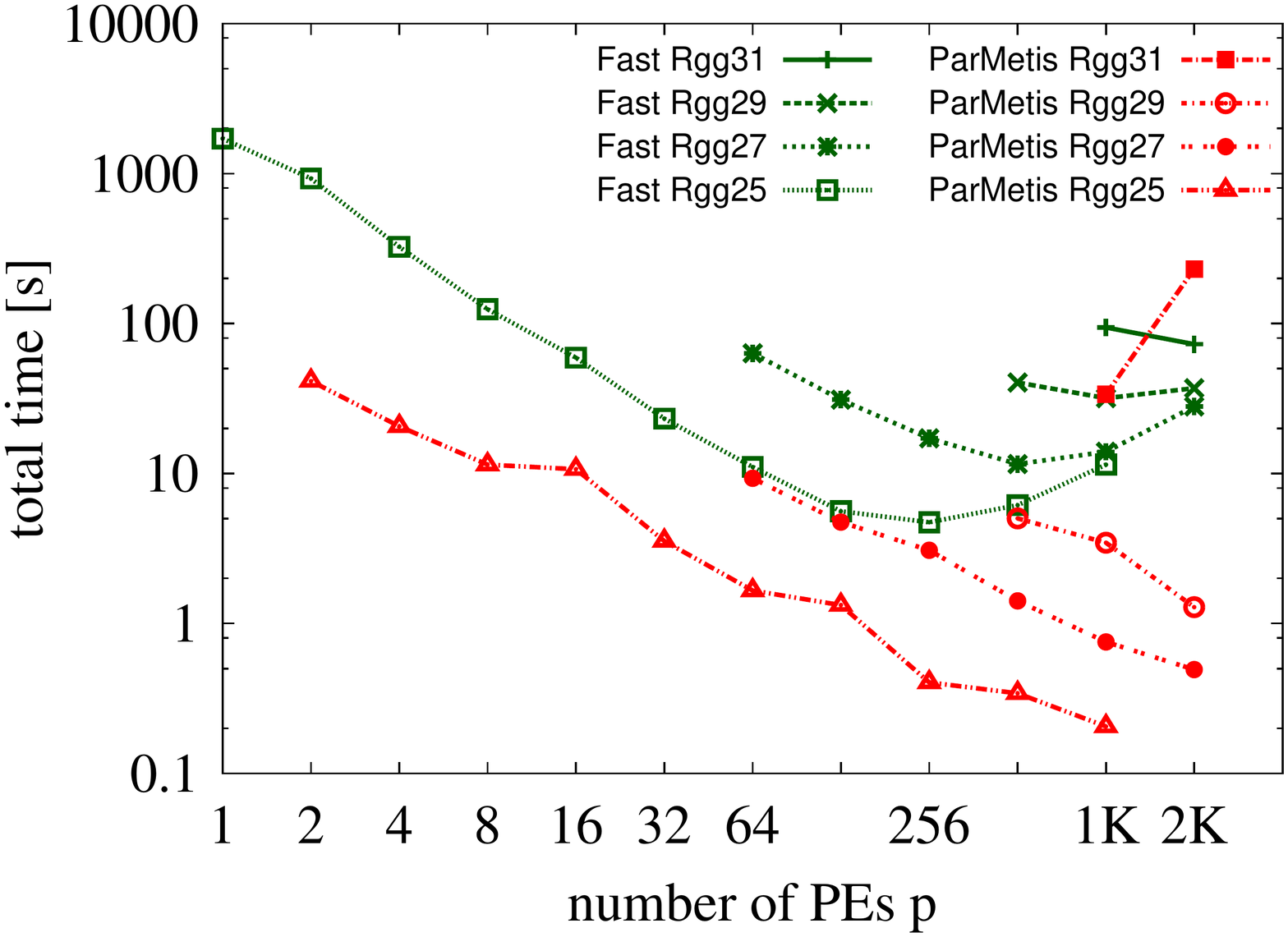}
        \vspace*{-.5cm}
        \hspace*{-.03\textwidth}
                \includegraphics[width=.5\textwidth]{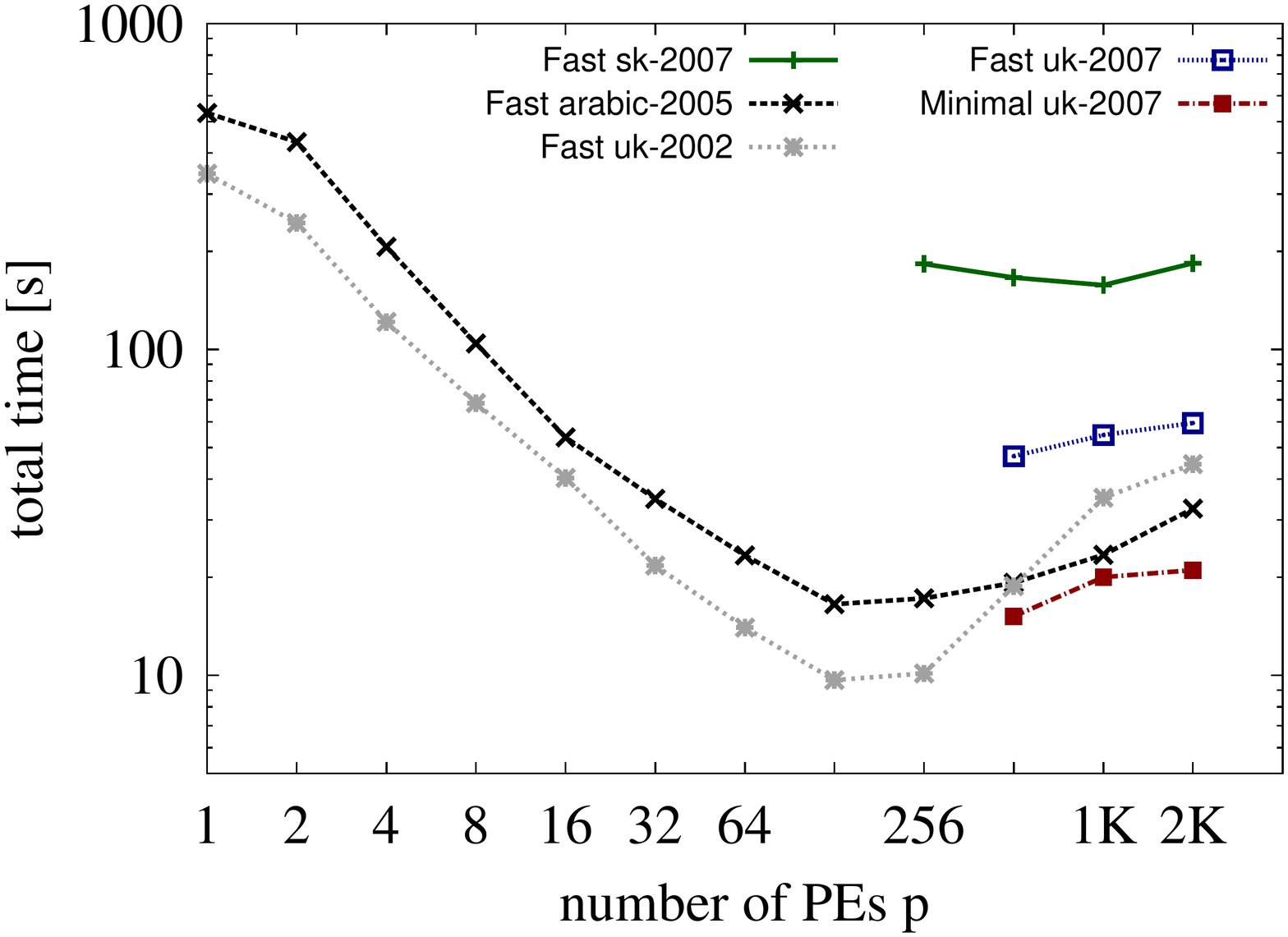}

        \caption{Top: Strong scaling experiments on Delaunay networks. The largest graph that ParMetis could partition from this graph family was \Id{del27}. Middle: Strong scaling experiments on random geometric networks. Bottom: Strong scaling experiments on the largest social networks from our benchmark set. Due to ineffective coarsening, ParMetis was not able to partition any of these graphs on machine B. On the largest graph, uk-2007, we also used the \Id{minimal} variant of our algorithm. \emph{Note}, although our system is not built for mesh-type networks such as Delaunay and random geometric graphs, we can partition larger instances and compute better solutions than ParMetis. 
        }
        \label{fig:strongscalingrgg}
\end{figure}

\label{s:expweakscalability}
We now look at \emph{strong scalability}. Here we use a subset of the random geometric and Delaunay graphs as well as the four large graphs arabic, uk-2002, sk-2007, and uk-2007. In all cases we use up to 2048 cores of machine B (except for del25 and rgg25, for which we only scaled up to 1024 cores).
Again, we focus on the \Id{fast} configuration of our algorithm and ParMetis to save running time. Figure~\ref{fig:strongscalingrgg} summarizes the results of the experiments.
First of all, we observe strong scalability to thousands of processors if the graphs are large enough. 
On del29 and rgg31, our algorithm scales all the way down to 2048 cores. 
Using all 2048 cores, we need roughly 6.5 minutes to partition del31 and 73 seconds to partition rgg31. 
Note that the rgg31 graph has three times more edges than del31 but the running time needed to partition del31 is higher. 
This is due to the fact that the Delaunay graphs have very bad locality, \ie when partitioning del31 more than 40\% of the edges are ghost edges, whereas we observe less than 0.5\% ghost edges when partitioning the largest random geometric graph.
Although the scaling behaviour of ParMetis is somewhat better on the random geometric graphs rgg25-29, our algorithm is eventually more than three times faster on the largest random geometric graph under consideration when all 2048 cores are used. 

As on machine A, ParMetis could not partition the instances  uk-2002, arabic, sk-2007 and uk-2007 -- this is again due to the amount of memory needed arising from ineffective coarsening.
On the smaller graphs, uk-2002 and arabic, our algorithm scales up to 128 cores obtaining a 35-fold and 32-fold speed-up compared to the case where our algorithm uses only one PE.
On the larger graphs sk-2007 and uk-2007 we need more memory. The smallest number of PEs needed to partition sk-2007 and uk-2007 on machine B has been 256 PEs and 512 PEs, respectively.
We observe scalability up to 1K cores on the graph sk-2007 (although, to be fair, the running time does not decrease much in that area). 
On uk-2007 we do not observe further scaling when switching from 512 - 2048 cores so that it is unclear where the sweet spot is for this graph. 
We also applied the \Id{minimal} configuration on machine B to the largest web graph uk-2007 in our benchmark set. The \Id{minimal} configuration needs 15.2 seconds to partition the graph when 512 cores are used. 
The cut is 18.2\% higher compared to the cut of the \Id{fast} configuration, which needs roughly 47 seconds to perform the partitioning task and cuts approximately 1.03M edges on average. This is fifty-seven times faster than the time needed to partition this graph using one core of machine A (which is faster).

\label{s:expstrongscalability}
\label{s:comparisonetc}

\section{Conclusion and Future Work}
\label{s:conclusion}
Current state-of-the-art graph partitioners have difficulties when
partitioning massive complex networks, at least partially due to ineffective coarsening.
We have demonstrated that high quality partitions of such networks can be obtained in
parallel in a scalable way. 
This was achieved by using a new multilevel scheme based on the contraction of size-constrained clusterings, which can reduce the size of the graph very fast.
The clusterings have been computed by a parallelization of the size-constrained label propagation algorithm \cite{pcomplexnetworksviacluster}.
As soon as the graph is small enough, we use a coarse-grained distributed memory parallel evolutionary algorithm to compute a high quality partitioning of the graph.
By using the size-constraint of the graph partitioning problem to solve, the parallel label propagation algorithm is also used as a very simple, yet effective, local search algorithm. Moreover, by integrating techniques like V-cycles and the evolutionary algorithm on the coarsest level, our system gives the user a gradual choice to trade solution quality for running time. 

The strengths of our new algorithm unfolds in particular on social networks and web graphs, where average solution quality \emph{and} running time is much better than what is observed by using ParMetis. Due to the ability to shrink highly complex networks drastically, our algorithm is able to compute high quality partitions of web scale networks in a matter of seconds, whereas ParMetis fails to compute any partition. 
Moreover, our algorithm scales well up to thousands of processors.

Considering the good results of our algorithm, we want to further improve and release it.
%
An important emerging application scenario are large-scale graph processing toolkits based on cloud computing.
Significant savings for several algorithmic kernels within the toolkit GPS have been reported by using 
graph partitioning~\cite{Salihoglu:2013:GGP:2484838.2484843} -- ParMetis in their case. Due to the superiority of our new algorithm compared to ParMetis on large
complex networks, further running time savings can be anticipated, also for related tools like 
Google's Pregel~\cite{Malewicz:2010:PSL:1807167.1807184}, Apache Giraph (\url{https://giraph.apache.org/}), Giraph++~\cite{tian2013think}, and GraphLab~\cite{Low:2012:DGF:2212351.2212354}. 
In future work we want to develop a very fast prepartitioner for such systems and we want to take advantage of the already computed partition in later multilevel iterations to further minimize the communication needed by the label propagation algorithm. 
Our algorithm may also be very helpful if a prepartition of the graph is already available, \eg from geographic initializations as in \cite{UganderB13}. 
This prepartition could be directly fed into the first V-cycle and consecutively be improved. 
In practical applications it may be advantageous to impart the information given by ground-truth communities if such are available.

It will be very interesting to generalize our algorithm for graph clustering w.r.t. modularity. 
For example, it should be straightforward to integrate the algorithm of Ovelgönne and Geyer-Schulz \cite{ogs12} to compute a high quality modularity graph clustering on the coarsest level of the hierarchy. 
This would enable researchers to compute graph clusterings of huge unstructured graphs in a short amount of time.
Another issue that we want to look at are other objective functions. 
For example, it might be interesting to integrate other objective functions such as maximum/total communication volume or maximum quotient graph degree into the evolutionary algorithm which is called on the coarsest graph of the hierarchy as well as into the label propagation algorithm. 

\section*{Acknowledgements}
We would like to thank the Steinbuch Centre of Computing for giving us access to the Instituts Cluster II. 
Moreover, we would like to thank Horst Gernert and his team for the support that we received while we performed the scalability experiments on the cluster.
\bibliographystyle{IEEEtran}
\bibliography{phdthesiscs,refs-parco}
\begin{appendix}
\begin{table*}[t]
\vspace*{-7.5cm}
\small
\centering
\caption{Average performance (cut and running time) and best result achieved by different partitioning algorithms. Results are for the case $k=32$. All tools used 32 PEs of machine A. Results indicated by a * mean that the amount of memory needed by the partitioner exceeded the amount of memory available on that machine when 32 PEs are used (512GB RAM). The ParMetis result on arabic has been obtained using 15 PEs (the largest number of PEs so that ParMetis could solve the instance).}
\label{tab:bipartitioningresults}
\begin{tabular}{|l||r|r|r||r|r|r||r|r|r|}
\hline
 algorithm  &  \multicolumn{3}{c||}{ParMetis} & \multicolumn{3}{c||}{Fast} &  \multicolumn{3}{c|}{Eco} \\
\hline
graph &    avg. cut & best cut & $t$[s]& avg. cut & best cut & $t$[s] &  avg. cut & best cut & $t$[s]\\
\hline
\hline
amazon      & \numprint{253568}   & \numprint{249071}            & \numprint{0.62}     & \numprint{235614}   & \numprint{231169}   & \numprint{3.20}   & \numprint{224550}   & \textbf{\numprint{222450}}  & \numprint{81.83}\\
eu-2005     & \numprint{974279}   & \textbf{\numprint{951537}}   & \numprint{33.28}    & \numprint{1218484}  & \numprint{1154916}  & \numprint{3.30}   & \numprint{1089613}  & \numprint{1010128}          & \numprint{79.87}\\
youtube     & \numprint{918520}   & \numprint{916657}            & \numprint{10.41}    & \numprint{951591}   & \numprint{936333}   & \numprint{13.86}  & \numprint{905330}   & \textbf{\numprint{889941}}  & \numprint{137.61}\\
in-2004     & \numprint{34445}    & \numprint{32711}             & \numprint{4.76}     & \numprint{26618}    & \numprint{25819}    & \numprint{1.97}   & \numprint{23795}    & \textbf{\numprint{22371}}   & \numprint{73.22}\\
packing     & \numprint{349000}   & \numprint{343611}            & \numprint{0.28}     & \numprint{338458}   & \numprint{335732}   & \numprint{24.65}  & \numprint{318242}   & \textbf{\numprint{315684}}  & \numprint{92.55}\\
enwiki      & \numprint{32539098} & \textbf{\numprint{32279759}} & \numprint{364.88}   & \numprint{33464700} & \numprint{33256794} & \numprint{787.41} & \numprint{33358352} & \numprint{32579351}         & \numprint{989.20}\\
channel     & \numprint{934264}   & \textbf{\numprint{919975}}   & \numprint{0.63}     & \numprint{989570}   & \numprint{983211}   & \numprint{26.23}  & \numprint{932175}   & \numprint{927128}           & \numprint{100.86}\\
hugebubbles & \numprint{28844}    & \numprint{28443}             & \numprint{4.98}     & \numprint{27832}    & \numprint{27607}    & \numprint{117.72} & \numprint{25358}    & \textbf{\numprint{25102}}   & \numprint{342.75}\\
nlpkkt240   & \numprint{7296962}  & \textbf{\numprint{7217145}}  & \numprint{17.09}    & \numprint{8048555}  & \numprint{7987330}  & \numprint{104.96} & \numprint{7770995}  & \numprint{7726512}          & \numprint{274.67}\\
uk-2002     & \numprint{2636838}  & \numprint{2603610}           & \numprint{193.48}   & \numprint{1710106}  & \numprint{1677872}  & \numprint{33.44}  & \numprint{1635757}  & \textbf{\numprint{1610979}} & \numprint{218.27}\\
del26       & \numprint{167208}   & \numprint{165361}            & \numprint{23.04}    & \numprint{153835}   & \numprint{152889}   & \numprint{274.75} & \numprint{145902}   & \textbf{\numprint{145191}}  & \numprint{859.33}\\
rgg26       & \numprint{423643}   & \numprint{419911}            & \numprint{8.14}     & \numprint{356589}   & \numprint{352749}   & \numprint{125.07} & \numprint{326743}   & \textbf{\numprint{323997}}  & \numprint{376.27}\\
arabic-2005 & *\numprint{4095660} & *\numprint{3993166}          & *\numprint{1414.83} & \numprint{3309602}  & \numprint{2648126}  & \numprint{45.40}  & \numprint{2372631}  & \textbf{\numprint{2178837}} & \numprint{251.38}\\
sk-2005     & *                   & *                            & *                   & \numprint{58107145} & \numprint{46972182} & \numprint{693.91} & \numprint{34858430} & \textbf{\numprint{29868523}}& \numprint{2183.10}\\
uk-2007     & *                   & *                            & *                   & \numprint{5682545}  & \numprint{5114349}  & \numprint{223.68} & \numprint{4952631}  & \textbf{\numprint{4779495}} & \numprint{794.87} \\
\hline
\end{tabular}

\end{table*}
\vfill

\end{appendix}
\end{document}